\documentclass[journal=jacsat,manuscript=article]{achemso}

\usepackage{chemformula} 
\usepackage[T1]{fontenc} 
\usepackage{graphicx,xcolor}
\usepackage{colortbl}
\usepackage{booktabs}
\usepackage{algorithm}
\usepackage[noend]{algpseudocode}
\usepackage{changepage}
\usepackage{amsmath,amsfonts,amssymb,subcaption,graphicx,caption}

\author{Gary A. Sevison}
\affiliation[University of Dayton]{Department of Electro-Optics and Photonics, University of Dayton, Dayton, OH, USA}
\alsoaffiliation{Air Force Research Laboratory, Sensors Directorate, Wright-Patterson AFB, Ohio, USA}
\email{sevisong1@udayton.edu}
\author{Shiva Farzinazar}
\affiliation[University of California, Irvine]{Department of Mechanical Engineering, University of California, Irvine, Irvine, CA, USA}
\author{Joshua A. Burrow}
\affiliation[University of Dayton]{Department of Electro-Optics and Photonics, University of Dayton, Dayton, OH, USA}
\author{Christopher Perez}
\author{Heungdong Kwon}
\affiliation[Stanford University]
{Department of Mechanical and Aerospace Engineering, Stanford University, Stanford, CA, USA}
\author{Jaeho Lee}
\affiliation[University of California, Irvine]{Department of Mechanical Engineering, University of California, Irvine, Irvine, CA, USA}
\author{Mehdi Asheghi}
\author{Kenneth E. Goodson}
\affiliation[Stanford University]
{Department of Mechanical and Aerospace Engineering, Stanford University, Stanford, CA, USA}
\author{Andrew Sarangan}
\affiliation[University of Dayton]{Department of Electro-Optics and Photonics, University of Dayton, Dayton, OH, USA}
\author{Joshua Hendrickson}
\affiliation[Air Force Research Laboratory]{Air Force Research Laboratory, Sensors Directorate, Wright-Patterson AFB, Ohio, USA}
\author{Imad Agha}
\affiliation[University of Dayton]{Department of Electro-Optics and Photonics, University of Dayton, Dayton, OH, USA}
\alsoaffiliation{Department of Physics, The University of Dayton, 300 College Park, Dayton, OH 45469, USA}
\title[An \textsf{achemso} demo]
  {Phase change dynamics and 2-dimensional 4-bit memory in Ge\textsubscript{2}Sb\textsubscript{2}Te\textsubscript{5} via telecom-band encoding}


\begin{document}



\begin{abstract}
As modern  computing gets continuously pushed up against the von Neumann Bottleneck -limiting the ultimate speeds for data transfer and computation- new computing methods are needed in order to bypass this issue and keep our computer's evolution moving forward, such as hybrid computing with an optical co-processor, all-optical computing, or photonic neuromorphic computing. In any of these protocols, we require an optical memory: either a multilevel/accumulator memory, or a computational memory. Here, we propose and demonstrate a 2-dimensional 4-bit fully optical non-volatile memory using Ge\textsubscript{2}Sb\textsubscript{2}Te\textsubscript{5} (GST) phase change materials, with encoding via a 1550 nm laser. Using the telecom-band laser, we are able to reach deeper into the material due to the low-loss nature of GST at this wavelength range, hence increasing the number of optical write/read levels compared to previous demonstrations, while simultaneously staying within acceptable read/write energies. We verify our design and experimental results via rigorous  numerical simulations based on finite element and nucleation theory, and we successfully write and read a string of characters using direct hexadecimal encoding.
\end{abstract}

\section{Introduction}

In recent years, computing systems have been approaching what is known as the von Neumann Bottleneck due to the inability of current processors to keep up with the needs of data transfer \cite{Kirchain2007,Alduino2007} to and from random access memories. One way to mitigate this issue is by moving towards hybrid systems with both electronic and optical components \cite{YZhao} or to other forms of computation that rely on non-traditional dynamic memories, such as accumulator  or computational memories \cite{Hosseini,Gao:2017} (that could be electronic or optical in nature). With the growing popularity of non-volatile phase change materials such as Germanium Antimony Telluride \cite{Cheng2018,Shu2013,Zalden2016} (and specifically the Ge\textsubscript{2}Sb\textsubscript{2}Te\textsubscript{5} (GST) stoichiometry) the possibility for a high-speed optical/electronic nonvolatile memory as a solution has gained traction \cite{Nirschl2007,Papandreou2010a,Papandreou2011,Gyanathan2011}. Phase change memories operate by storing the information in the binary change of material properties (e.g, refractive index, reflectivity, or electrical conductivity) that accompanies a transition between an amorphous state (RESET) and a crystalline state (SET).  One benefit of using GST is the ability of the material to be placed in many intermediate states of crystallization leading to a multilevel storage or accumulation. This has been shown recently to work using a waveguide structure with a strip of GST laying across the top \cite{Li2019,Rios2018,Zheng2018}. However, that geometry requires several fabrication steps and, due to the nature of the single-layer guided wave geometry, it limits the amount of real estate on chip that is available for storing data. Other geometries utilized a layered structure with top and bottom electrodes that can electronically switch the GST \cite{Joshi2011,Bedeschi2008}. These devices can be very small and compact (on the order of 10s of nms) \cite{Close2010,H-SPhilipWong2010}, but most are designed to be written and read electronically (hence not conducive to hybrid or all-optical computing)   and have the additional problem of the impedance of the device changing depending on the phase of the GST \cite{Guo2018}.
Simply being able to encode the data as in an optical platform is not enough to identify whether GST-based PCMs are  a viable option for a high bandwidth, dynamic, multilevel optical memory. The optical and thermal dynamics and time constants of the phase change need to be identified as well, especially if high-speed operation and long-term stability are desired. By understanding what happens during the crystallization of the material we can avoid or at least prolong the occurrence of issues that cause breakdown in the GST. One of the core issues, for example, is the Tellurium drift \cite{Kim2009,Yang2009,Kim2014} that occurs during melting and is accelerated by applying an electric field across the material. By optically switching the GST, as opposed to electrically switching it, and staying below the melting point while crystallizing, we can lengthen the lifetime of the material. The time dynamics of the change are additionally important to know in order to be able to obtain a limit to the eventual speed of any future device. Switching speeds for GST have been reported from the single nanoseconds or shorter \cite{Wuttig2012} to the hundreds of nanoseconds or microseconds \cite{Liang2011,Wright2000}, depending on the method used for the change as well as the power and wavelength of light for optical switching \cite{Zhai2019}. Similar measurements to those utilized in this paper have been used previously \cite{Zhang2012,Zhai2019,Khulbe2000,Liang2011}, however, no data was found for 1550 nm light, which is necessary to understand in order to create devices that work in the telecommunications bands. In fact, current free space optically induced phase change devices are operated in the visible or near-visible wavelength regime \cite{Khulbe2000,Lee2014,Waldecker2015,Simpson2015}. Since the absorption is extremely high at these wavelengths, a transmissive device is not a viable option. Moreover, the high absorption also limits the volume of material that can be changed via a free-space laser and therefore restricts the amount of levels that can be achieved. Hence, by using a telecom wavelength stimulus to trigger the phase change, where GST is inherently less absorbing,   we can gain access to more of the volume of the material to change. 

To sum up,  there is a real need for an optical memory to drive hybrid computing protocol; this memory has to a) be capable of high speed dynamic storage and retrieval, b) be capable of multilevel storage and accumulation, c) offer long-term stability and d) have reasonable power requirements. Moreover, in order to improve the performances in the future, there is a need for a deeper understanding of the material and device properties through both modeling and experimental verification.

 In this work, we present a 2D-ready memory that can surpass recent waveguide-based memories in terms of information density and can integrate with future optical computing protocols \cite{Xiang2018} that require 2D storage or a non-volatile accumulator. While we demonstrate single-pixel performance, we anticipate that our results can be easily scaled into a two-dimensional high-density device. Specifically, we demonstrate the use of  GST as a multilevel memory by encoding and decoding information based on a 4-bit all optical write/read scheme. Increasing numbers of levels is achieved by optimizing the GST thin-film thickness on silicon as well as the power in each pulse. Multiple levels are achieved by exciting thin films of GST with sequential single-pulses (ns) of 1550 nm light using a custom built optical setup capable of active optical monitoring. The crystallization time-dynamics are investigated experimentally and corroborated with numerical simulations revealing the partial crystallization increasing in a volume of GST with each subsequent pulse, with a goal of further improving GST-based memory devices in the future. 

\section{Opto-electrical modeling of 2-dimensional 4-bit GST based memories}
\label{sec:sim}


GST PCMs contain a rich set of coupled electrical, thermal, and optical properties. In order to begin studying these properties, numerical simulations were run for single and multi-pulse excitation with a focused, pulsed, telecom-band Gaussian beam. Once we begin to understand what was happening physically from the simulations it is possible to understand the optimal conditions needed to implement a high speed multilevel optical memory.

\begin{figure*}[htbp]
\centering
\begin{subfigure}{0.47\linewidth}
\centering
\includegraphics[width=\linewidth]{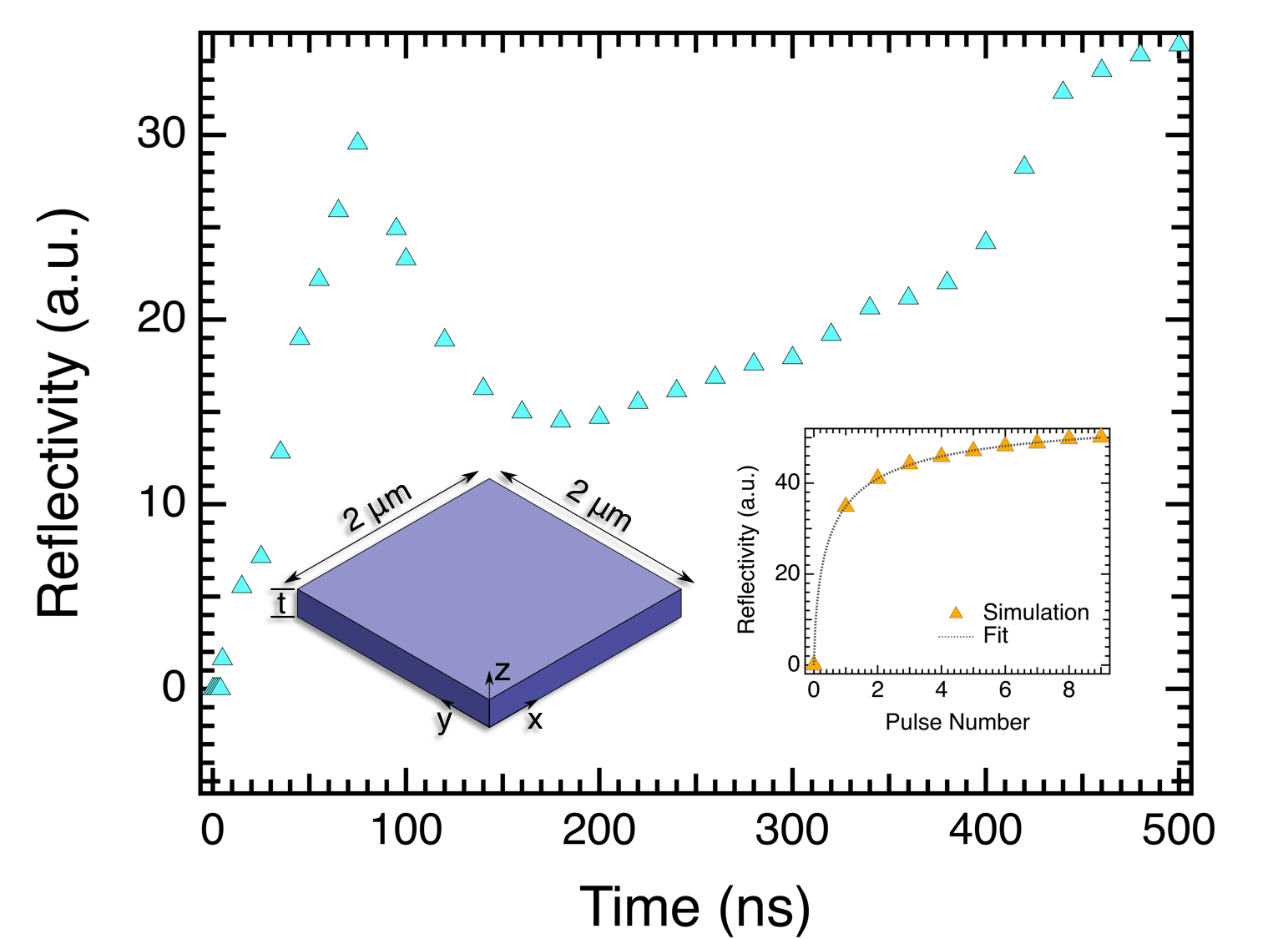}
\captionsetup{justification=centering}
\caption{}
\label{fig:Sim_SP}
\end{subfigure}
\begin{subfigure}{0.47\linewidth}
\centering
\includegraphics[width=\linewidth]{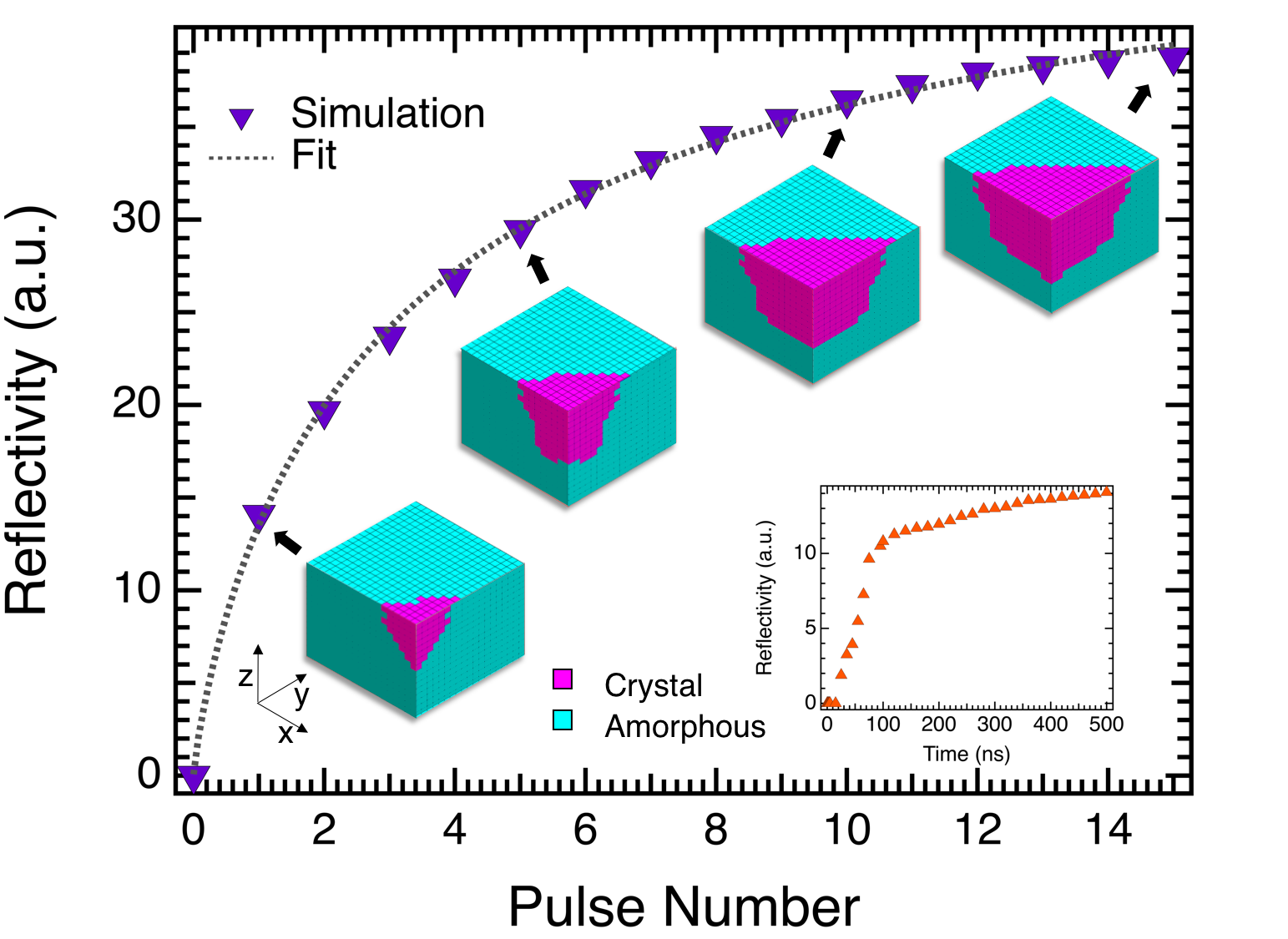}
\captionsetup{justification=centering}
\caption{}
\label{fig:Sim_MP}
\end{subfigure}
\caption{a) The temporal change dynamics of GST for a single 870 mW peak power pulse with 50 ns width and 25 ns falling tail. The left inset illustrates the simulation domain of the GST layer with t=530 nm in (a) and t=255 nm for the data in (b). The right inset in (a) shows the reflectivity change as a function of number of pulses for the pulse characteristics in (a). b) Reflectivity evolution as a function of pulse number for 350 mW peak power pulse with 50 ns width and 25 ns falling tail. Additionally, the images beneath the pulses shows a representation of the simulation domain created from the xy and xz planes with the xz plane being reflected to create a 3D image for ease of understanding. The lower right inset shows the temporal change dynamics for a single pulse with the same pulse dynamics as (b).}
\label{fig:sims}
\end{figure*}

In our simulations, we considered a volume of amorphous GST subjected to a sharply focused gaussian beam with a diameter of 4.00$\pm$0.25 $\mu$m, which will reflect our experimental conditions as seen later. For the reflectivity calculation, we used a fully coupled optothermal simulation domain to estimate the temperature evolution of the GST phases under a pulse with 50 ns width and 25 ns falling tail. Crystallization kinetics of the amorphous phase and solidification of the molten areas were assessed using the crystallization and nucleation model for phase transition as a function of temperature and time duration \cite{Peng1997}. $P_n$, the probability of amorphous phase transformation to a crystalline nucleus during the $\Delta t$ time interval, is defined as

\begin{equation}
P_n=\alpha _n \Delta t e^{- \beta \left [E_a+\frac{A}{(\Delta G)^2}\right ]}\text{,}
\label{eq:probn}
\end{equation}
where $E_a$ is the nucleation based activation energy and $\Delta G$ is the additional Gibbs free energy of amorphous GST. $A$ is associated with the interfacial surface free energy between crystalline and amorphous phases, $\alpha _n$ is the nucleation frequency factor, and $\beta$ is defined as $\frac{1}{k_B T}$ where $k_B$ is the Boltzmann constant and $T$ is the absolute temperature. Additionally, the probability of crystal growth ($P_g$) can be expressed as a function of the mesh size ($R$), activation energy associated with atomic diffusion ($E_{a2}$), growth frequency factor ($\alpha _g$), atomic jump distance ($\alpha _0$), and the melting temperature ($T_m$),

\begin{equation}
P_g=\Delta t \alpha _0 \alpha _g e^{\frac{-0.8}{\left (1-\frac{T}{T_m}\right )}}\left [1-e^{-\beta \Delta G}\right ]\frac{e^{-\beta E_{a2}}}{R}.
\label{eq:probg}
\end{equation}

The GST crystallization algorithm utilizing the two parallel mechanisms of probabilistic nucleus formation and crystalline growth was initially developed by Li et al. \cite{Li2012} for electrothermal modeling of GST based memories which we modified for optically actuated memories. The geometry and material properties were defined inside the commercial software COMSOL, which computed the temperature field at each time step. Phase change algorithms and optical switching were implemented using the MATLAB codes and properties were modified for the next iteration. The symmetry of the sample reduces the computation time by using only a quarter of the sample domain. The simulation considers 0.56 $\frac{W}{mK}$ and 0.37 $\frac{W}{mK}$  at room temperature for the cross-plane and in-plane crystalline thermal conductivity values and 0.18 $\frac{W}{mK}$ for the thermal conductivity of the amorphous phase \cite{Li2011,Lee2011,Lee2012}. A thermal boundary resistance of 0.9 $\frac{m^2KG}{W}$ was assumed between the GST and the silicon layer \cite{Lee2013}.  Further, the crystallite sizes of 14 nm and 27 nm for the face-centered cubic (fcc) and haxagonal close-packed (hcp) crystalline phases were estimated from the full width at half maximum approach on the XRD peaks of annealed GST at 170 °C and 350 °C, respectively. Using the local refractive indices at 1550 nm wavelength \cite{Guo2019} as an input into the frequency-dependent electromagnetic wave simulation domain, the reflectivity of the 255 nm thick GST film was calculated to be 0.547 and 0.433 in the crystalline and amorphous phases, respectively. For the 530 nm thick GST film these numbers became 0.555 and 0.341. Figure \ref{fig:Sim_SP} shows the simulated reflectivity as a function of time during the first applied pulse of 870 mW peak power incident on a 530 nm thick layer of GST. The initial reflectivity boost coincides with the ON state of the laser while the quench melting of the molten parts causes a decrease in reflectivity followed by a later growth due to the thermal diffusion and crystallization within the medium. Figure \ref{fig:Sim_MP} shows the final reflectivity at each pulse as a function of pulse number for a 350 mW peak power pulse incident on a 255 nm thick layer of GST.  As illustrated, the increase in the thermal conductivity driven by the crystallization phenomenon decreases the maximum temperature reached during each pulse; therefore, the reflectivity increase saturates at later pulses due to the annihilation of crystal growth. The saturation happens at earlier pulse numbers with higher pulse power as seen in the right inset in Fig. \ref{fig:Sim_SP} compared to Fig. \ref{fig:Sim_MP}. This saturation is an issue for reading back the levels written into the GST memory, however in section \ref{sec:memory} we discuss a process to minimize the impact on the final data. 

\section{Sample Preparation and Pump-Probe Setup}
\label{sec:setup}

In order to experimentally verify the numerical findings in section \ref{sec:sim}, we prepared GST thin films via magnetron sputtering from a GST ceramic target, indium bonded to a copper backing plate. Prior to deposition, the Si (100) substrates are pre-treated with a standard acetone-methanol-isopropanol rinse and a nitrogen air dry to clean the surface. A Si substrate is subsequently loaded into the sputtering chamber and pumped down to sub 1 $\mu$Torr pressure to eliminate unwanted particles and other polluting gases which may negatively affect the quality of the film. During the sputtering process, a 13.56 MHz radio frequency signal is applied to a 3 inch circular magnetron cathode with a power of 100 W and the working pressure is increased to 4 mTorr using Ar as the inert process gas. To achieve adequate thin film uniformity within 2 nm across a 3 inch wafer, the sample was rotated during the sputtering process.  We produced GST films on Si of thicknesses 255, 440, 530 and 1080 nm. We settled on using the 255 nm sample because it gave the clearest separation of the levels for the memory. Since this investigation only studies the crystallization process, we did not include a capping layer on the GST which is useful in preventing tellurium evaporation while the sample is in a molten state.

\begin{figure}[htbp]
\centering
\includegraphics[width=\linewidth]{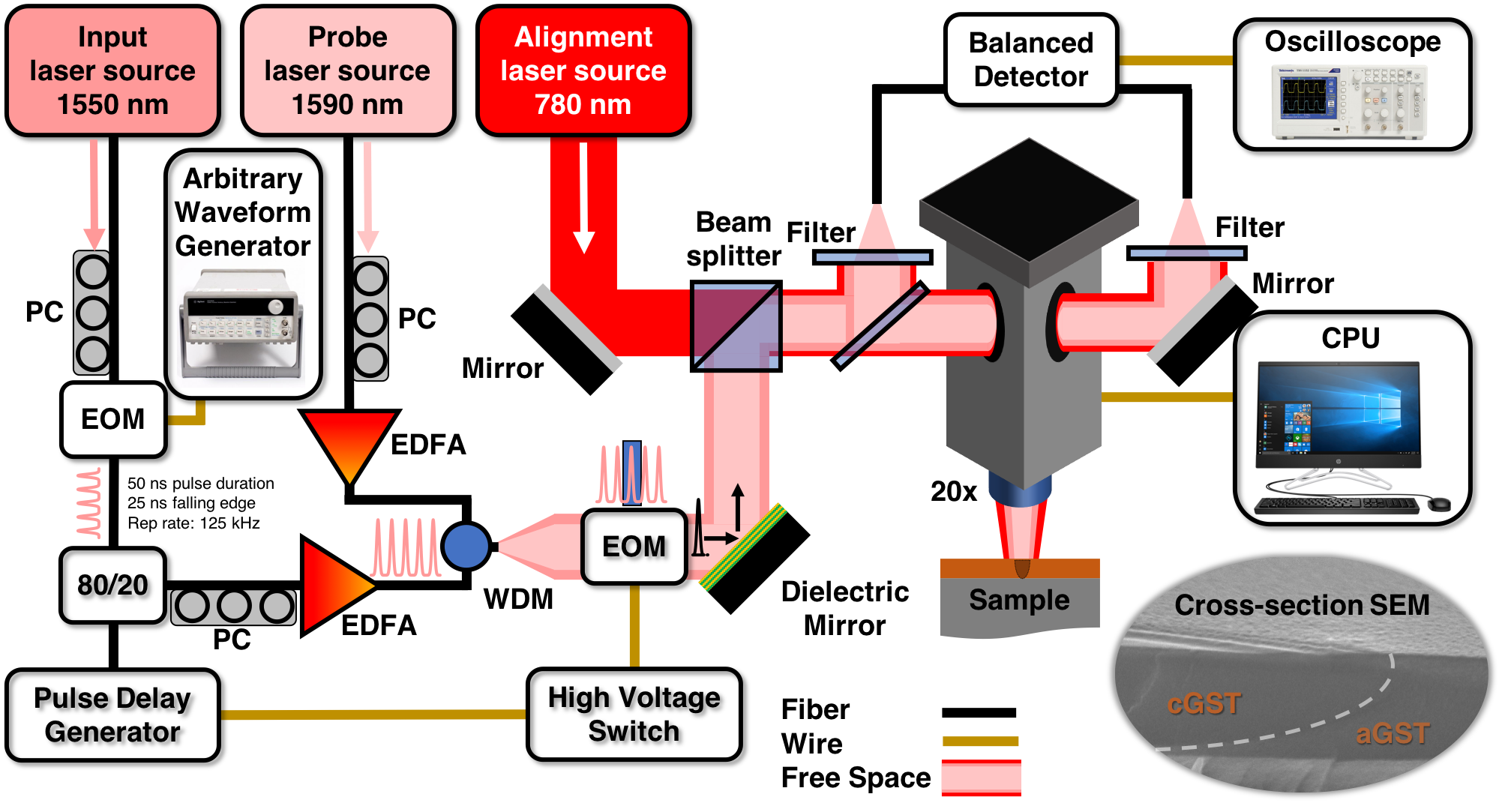}
\caption{Single shot pump-probe measurement setup.}
\label{fig:setup}
\end{figure}

After fabrication, the samples were actuated and monitored in a custom built optical experimental setup, shown in Fig. \ref{fig:setup}, capable of producing single pulses down to 25 ns and up to constant wave (CW) using 1550 nm light. The setup begins with a 1550 nm CW laser that runs through a fiber coupled electro-optic modulator (EOM) connected to an arbitrary waveform generator (AWG) which allows us to create any pulse shapes that may be needed. We then use an Erbium doped fiber amplifier (EDFA) to amplify the power in the pulses to powers adequate for switching the GST. At this point we utilize a wavelength demultiplexer (WDM) to couple our probe beam, which is a CW 1590 nm laser, into the same fiber via the second port. This co-aligns the lasers before they leave the fiber saving alignment time and loss from additional optics. The now combined beams then exit the fiber and pass through a free space EOM which can handle the higher powers of our pulses. The EOM is used to pick off as many pulses as we need in a single shot. As we were unable to find a high voltage driver for the EOM that could provide the voltage needed with a fast enough slew rate to pick off a single pulse, we created our own pulse amplifier circuit using a high-slew rate operational amplifier (APEX Microdevices PA94). We achieved a 400 V jump in less than 5 $\mu$s for an arbitrary wave shape. This allowed us to easily capture a single pulse from a train of pulses up to few hundred kHz. After the free-space EOM, we co-align a 780 nm laser to use for positioning on the sample since the 1550 and 1590 nm wavelengths we are working with are not visible on our CMOS camera.

\begin{figure}[htbp]
\centering
\includegraphics[width=0.7\linewidth]{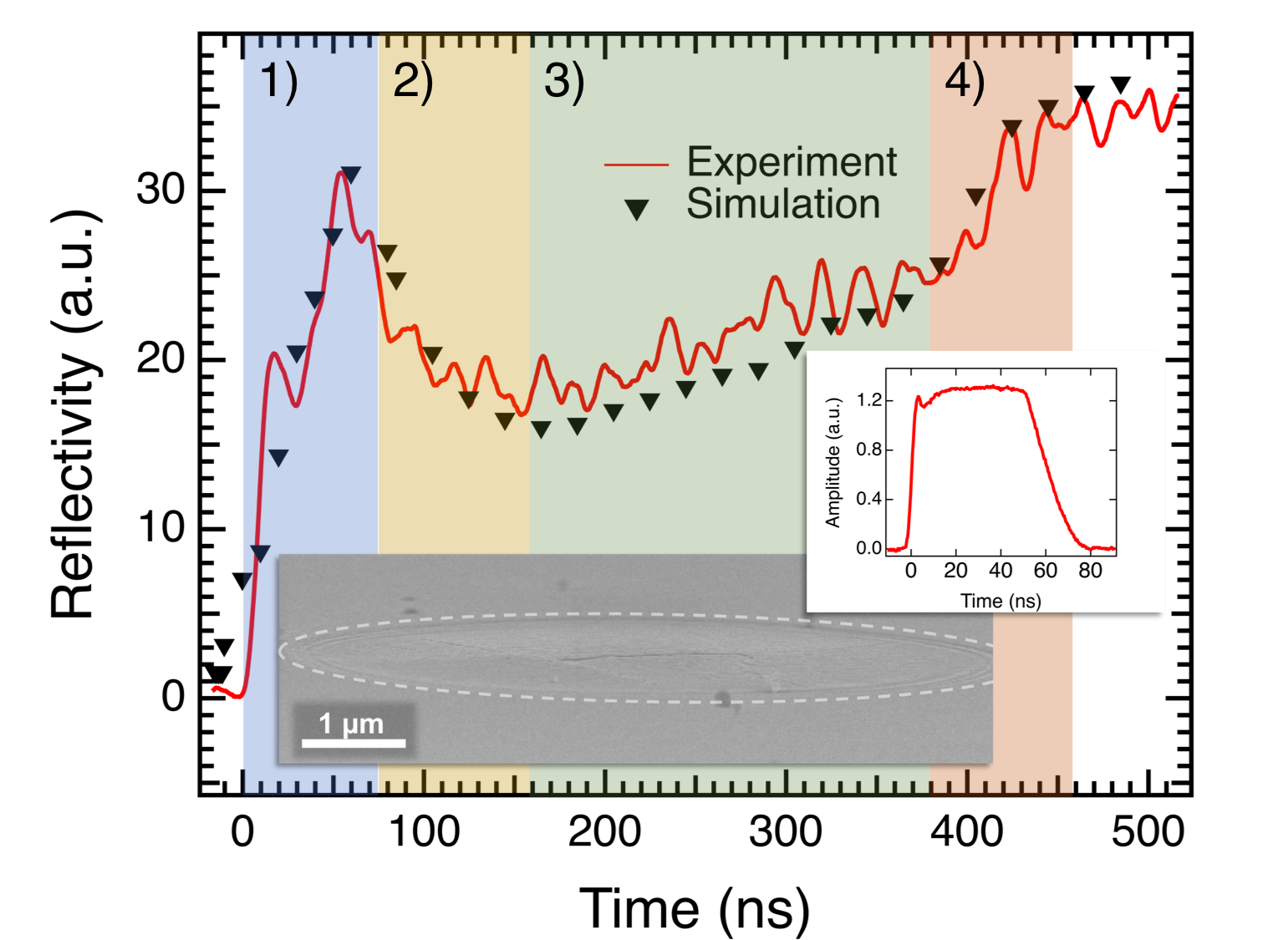}
\caption{Single pulse (870 mW peak power, 50 ns pulse, 25 ns tail) time dynamics from 530 nm thick GST sample. Colored areas label the different stages of crystallization. Top inset shows 50 ns pulse with 25 ns tail. Bottom inset is an SEM image of GST illuminated by a single pulse.}
\label{fig:singlea}
\end{figure}

After the EOM, a glass slide is used to pick off a small portion of the beam. This becomes one arm of the balanced detection scheme and is also used as a power monitor. The beam then passes through a 1590 nm bandpass filter, in order to filter out the 1550 nm pump pulse as much as possible, and is then coupled into a fiber. The coupling efficiency into the fiber is used to match the powers before and after the sample for the balanced measurement. The majority of the beam (~90\%) enters the microscope and is focused down to a 4-5 $\mu$m spot on the sample. After reflecting off the sample the beam is picked up from the microscope's output and passes through another 1590 nm bandpass filter before being coupled into a fiber and inserted into the detector as the second arm of the balanced detection scheme. The detector then subtracts the two signals and outputs the final signal to an oscilloscope where it is captured.

\section{Phase change dynamics}
\label{sec:dynamics}

Using the pump-probe setup, we are able to observe the temporal dynamics of the phase change of the GST via reflection measurements. While the absolute reflectivity contrast is appreciable, losses in the microscope - which was initially designed for visible epi-fluorescence -  meant that the reflected light was very low power. The balanced measurement aspect of our setup, allowed us to improve the signal-to-noise ratio in our reflected probe, giving a better dynamic range for the measurement and a more full understanding of the processes occurring.

Figure \ref{fig:singlea} shows the reflectivity measurement for a single 870 mW peak power pulse incident on our sample. Four main areas are highlighted indicating the different parts of the process that occur during the crystallization of GST by a single 50 ns pulse with a 25 ns falling tail and experimentally verify what was seen in the numerical simulations from Fig. \ref{fig:Sim_SP}. Taking into account the differing pulse lengths, the crystallization times did not differ significantly for 25-100 ns pulses and were therefore not shown here. During the initial illumination by the pulse, nucleation sites are created in the GST. These sites act as the seeds for the crystallization process. From these nucleation sites being created we observe a fast rise in the reflectivity of the material. This is highlighted in the first shaded area marked as 1) in Fig. \ref{fig:singlea}.


After the pulse ends the energy diffuses through the GST. For higher powered pulses a section of the data can be seen to drop in reflectivity, e.g. the second shaded area labeled 2) and at the end of area 1) in Fig. \ref{fig:singlea}. This is due to the GST reaching its melting temperature \cite{Khulbe2000}. As the material melts it approaches its amorphous phase and shows a decrease in reflectivity, just as seen here.

\begin{figure}[htbp]
\centering 
\includegraphics[width=0.7\linewidth]{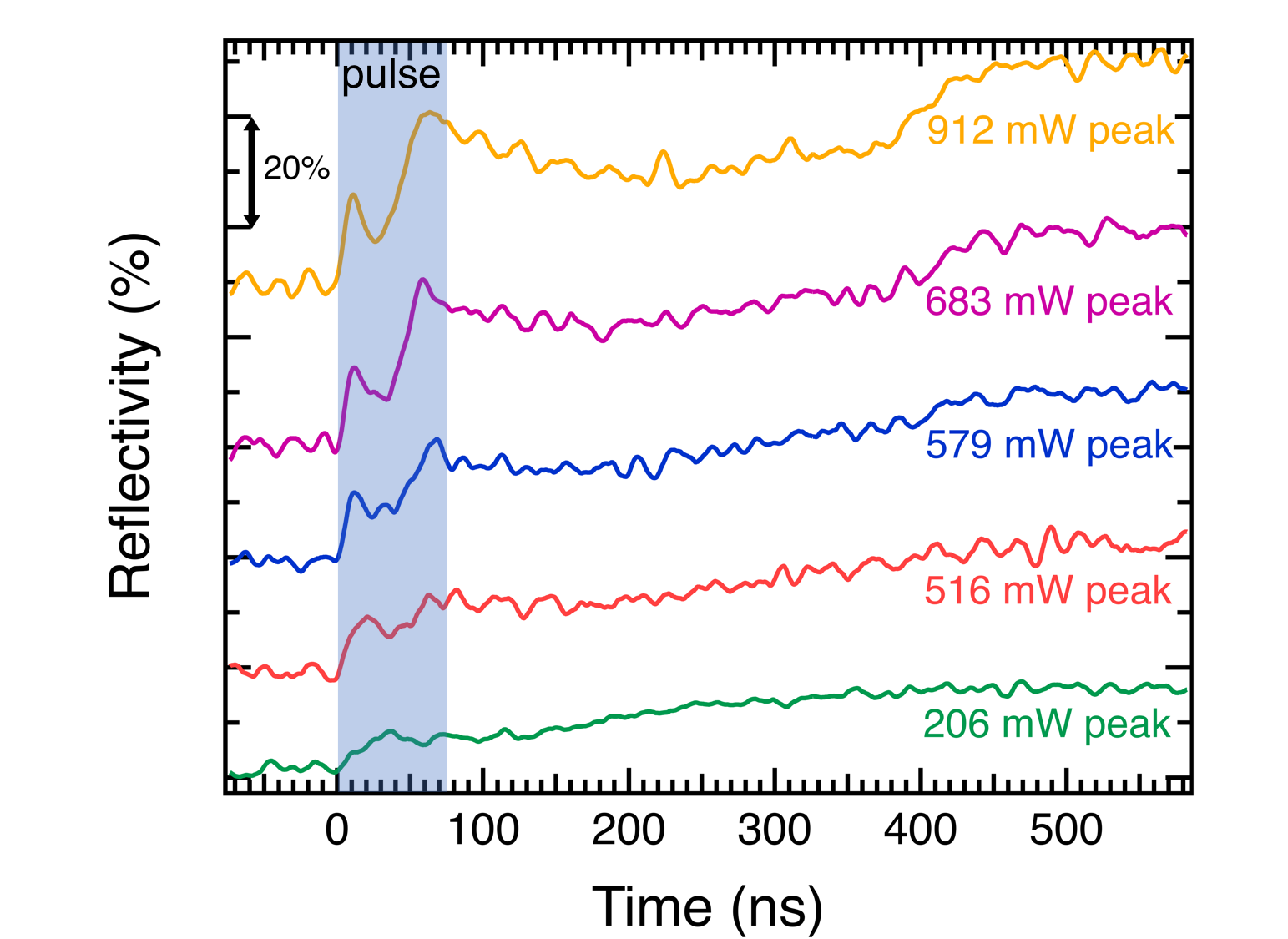}
\caption{Single pulse dynamics for differing pulse powers. Traces are shifted vertically for clarity.}
\label{fig:singleb}
\end{figure}

As the heat continues to diffuse through the material the temperature drops below the melting point and we begin seeing an increase in reflectivity due to the crystallization occurring. The shaded region 3) shows this happening with a slowly rising reflectivity. The slow rise is due to the fact that the GST is still above its crystallization temperature and therefore still has the energy to break the bonds needed for crystallization. Once the material cools below the crystallization temperature a rapid increase in reflectivity is seen in the shaded area 4). The crystallization is no longer being impeded by the energy from the pulse and can relax into its final crystallized state.

Since melting the GST allows the Tellurium atoms to drift more easily, which is one of the leading causes of switching failure \cite{Kim2009,Yang2009,Kim2014}, staying away from melting while crystallizing is crucial. In Fig. \ref{fig:singleb} it can be seen that properly selecting the correct power for the pulses can keep the material from melting. This phenomena can be seen in the 206 mW peak power trace in the figure. From Fig. \ref{fig:singleb} we can also see that the GST crystallizes more with higher pulse powers. However there is a limit to this which is seen between the 683 mW and 912 mW peak power pulses. There is not a significant difference in the final reflection levels of these two pulses, but the 912 mW peak power trace dips lower after the initial height after the pulse ends, indicating more melting.

\begin{figure}[htbp]
\centering 
\includegraphics[width=0.7\linewidth]{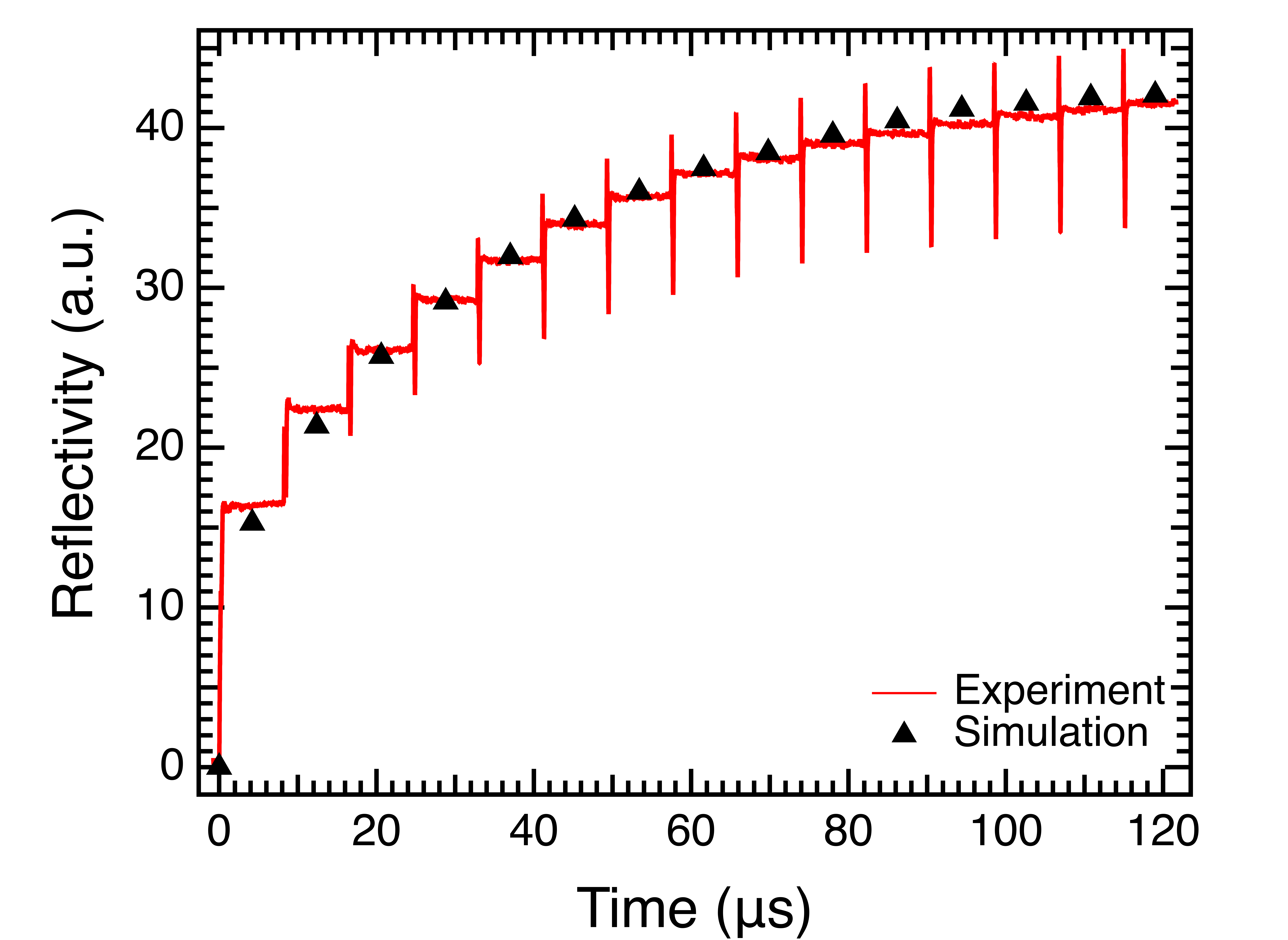}
\caption{Simulation and experimental data for $\sim$350 mW pulses. Each point of the simulated data and each step of the experimental data represents an incremental number of pulses incident on the sample.}
\label{fig:multi}
\end{figure}

In Fig. \ref{fig:multi} we show the experimental and simulated data for 15 $\sim$350 mW pulses incident on a 255 nm film of GST. Each step in the data signifies another pulse incident on the sample. The peaks seen before the reflectivity settles down to its steady state are leakage from the pump laser through the filters. As more of the GST is switched the reflectivity goes up and more of the pump is leaked. As can be seen in Figs. \ref{fig:singlea} and \ref{fig:multi}, the simulated data matches the experimental data very closely. By reducing the power of the single pulses a clearer multi level memory can be achieved. We demonstrate this in the next section where we create an optical 4-bit memory.

\section{Multilevel Memory}
\label{sec:memory}

\begin{figure*}[htbp]
\centering
\begin{subfigure}{\linewidth}
\centering
\includegraphics[width=\linewidth]{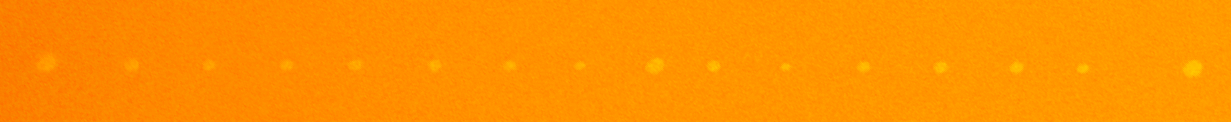}
\captionsetup{justification=centering}
\caption{}
\label{fig:write}
\end{subfigure}\\
\begin{subfigure}{0.47\linewidth}
\centering
\includegraphics[width=\linewidth]{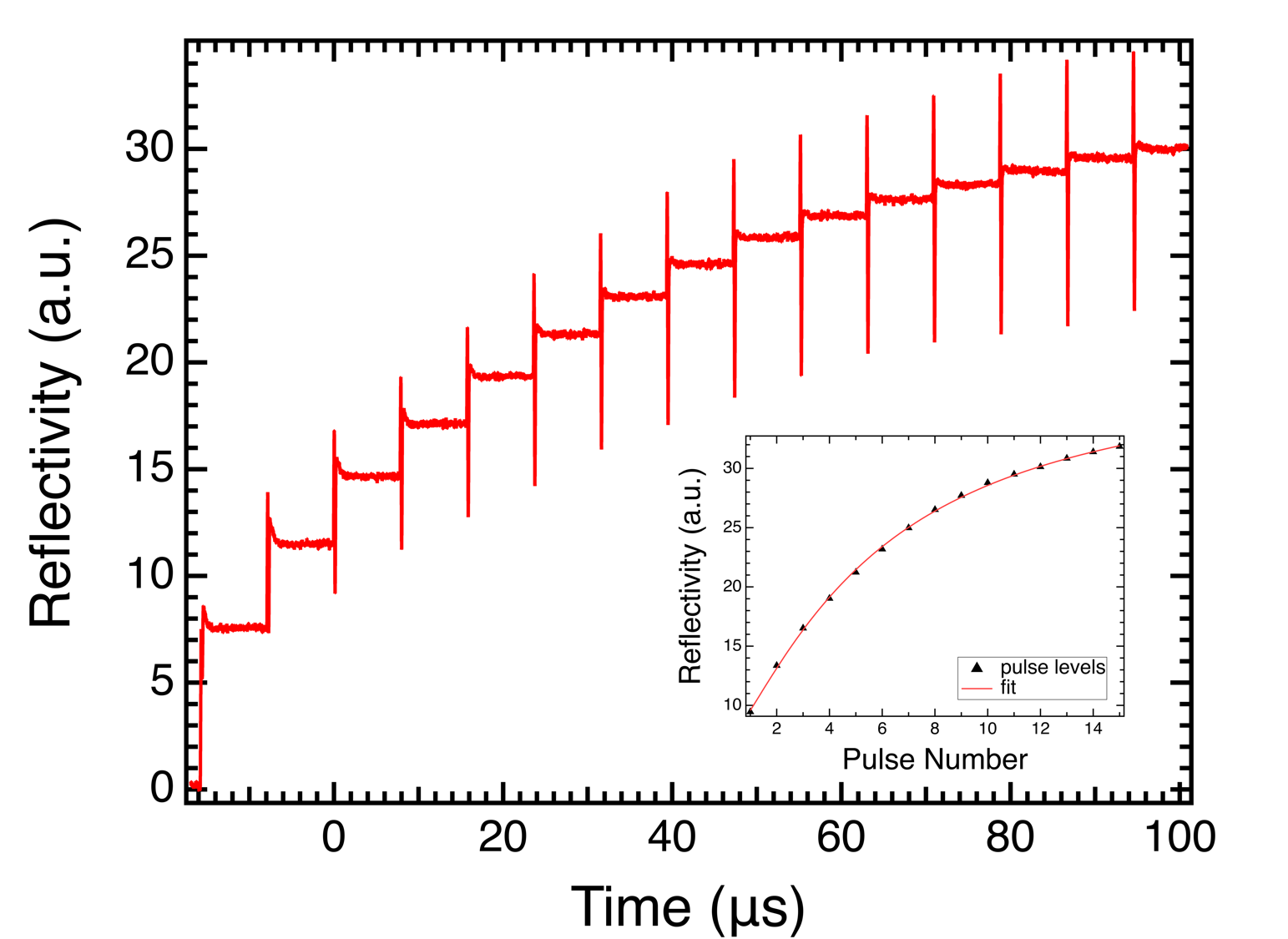}
\captionsetup{justification=centering}
\caption{}
\label{fig:16level}
\end{subfigure}
\begin{subfigure}{0.47\linewidth}
\centering
\includegraphics[width=\linewidth]{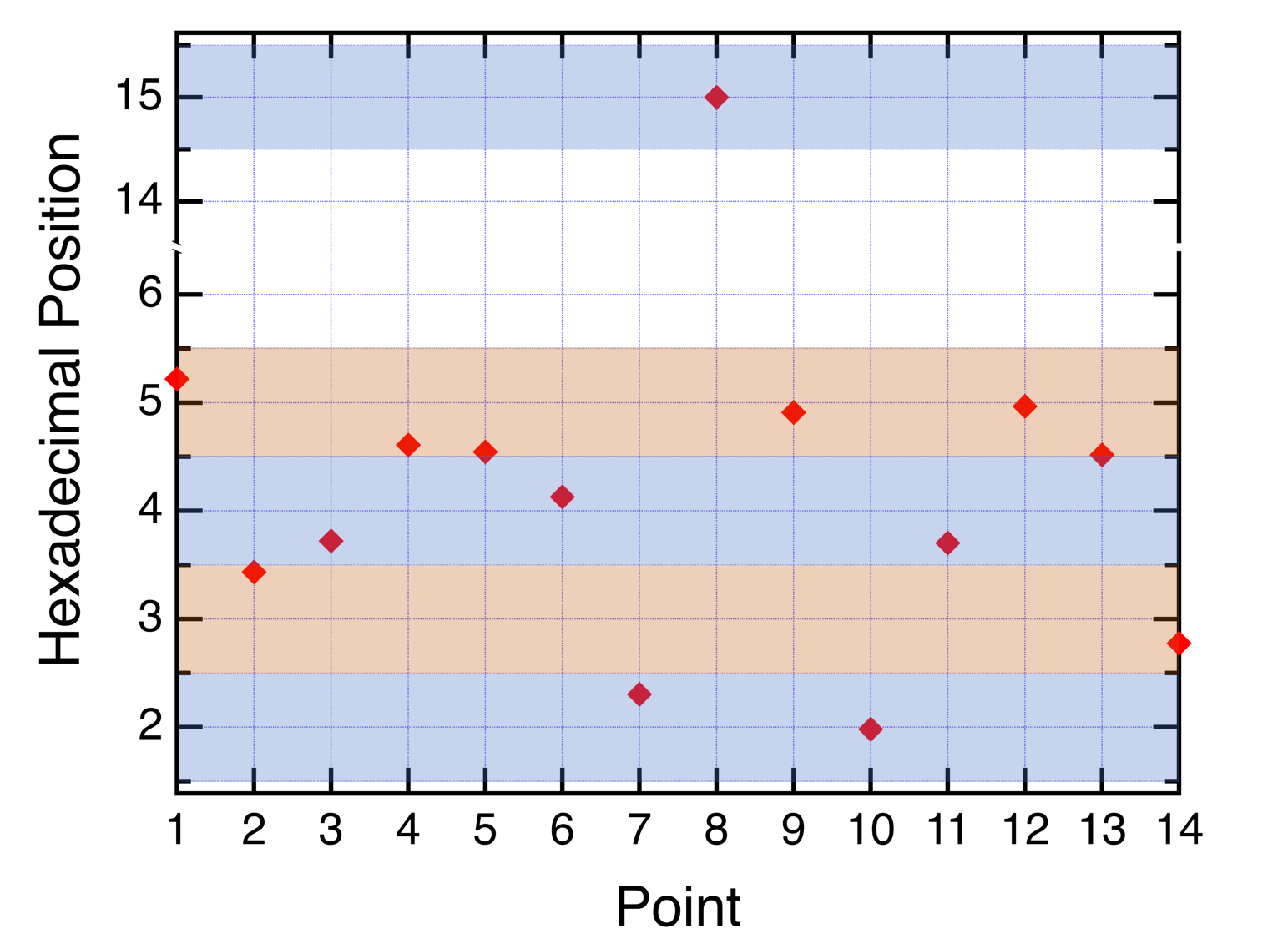}
\captionsetup{justification=centering}
\caption{}
\label{fig:final}
\end{subfigure}
\caption{a) "SET/RES" encoded in hexadecimal in GST film. The first and last pulse are 16 pulses used as markers for the beginning and ending of the message. b) 16 levels achieved using multiple plulses with peak power of $\sim$200 mW. Inset shows the fit to the levels using the hill equation \ref{eq:HillEquation}. c) Final read data from "SET/RES".}
\label{fig:readwrite}
\end{figure*}

With the knowledge that different amounts of crystallization occur at different pulse powers, we attempted to use multiple pulses to achieve these levels, which could be looked at as either a single-site multi-bit storage or an accumulator memory (since the low repetition-rate implies the material reaches steady-state between the pulses) . Figure \ref{fig:16level} shows an example of a sixteen level write with some smoothing used to differentiate the levels more easily. For the smoothing a box algorithm was employed using 500 points. This figure was obtained by monitoring the reflection off the sample as a single spot was illuminated with 15 pulses of peak power $\sim$ 200 mW. It can be seen that there exists a saturation as higher pulse numbers were incident on the GST. This needs to be corrected for in order to read data encoded with these pulse powers. 

To correct for the saturation curve the individual levels were pulled from the graph in Fig. \ref{fig:16level} and were fit using the Hill Equation 

\begin{equation}
y=\frac{base+(max-base)}{1+\left [\frac{x_{1/2}}{x}\right ]^{rate}},
\label{eq:HillEquation}
\end{equation}
as seen in the inset in Fig. \ref{fig:16level}, where $base$ sets the $y$ value at small $x$, $max$ sets the $y$ value at large $x$, and $x_{1/2}$ is the $x$ value where $y=\frac{base+max}{2}$. Once the fit parameters were known we could solve the equation for $x$ and use

\begin{equation}
x=\frac{x_{1/2}}{\left [\frac{max-base}{y-base}-1\right ]^{\frac{1}{rate}}}
\label{eq:HillEquation_solved}
\end{equation}
to decode our data.

Using the same power and size of pulses we then encoded an initial 16 pulse spot and a final 16 pulse spot at the two ends of where the message would be written. The space between these two spots was then scanned and acted as the baseline for reading the peaks after they were written. After reading the baseline, "SET/RES" was written into the GST by stepping the sample under the laser by a fixed distance and hitting each spot with the corresponding number of pulses for hexadecimal encoding, "53 45 54 2f 52 45 53". Where the "f" corresponds to the 15th level. The resulting image of the spots, taken by a visible wavelength microscope, can be seen in Fig. \ref{fig:write}.


The encoded data was then scanned with the 1590 nm CW probe laser and corrected via background scans of the same area on the sample. The peak data was taken from each of the written spots, run through the decoding algorithm in equation \ref{eq:HillEquation_solved}, and graphed as shown in Fig. \ref{fig:final}. The highest value was set to the $15^{th}$ level, as that is the maximum level of our memory.



As can be seen in Fig. \ref{fig:final}, the message "SET/RES" was able to be read back within the expected values $\pm 0.5$. With a more consistent laser and a more efficient probing setup, this could be extended to higher levels and even smaller margins of error.


Since our memory device is encoded using a laser incident on the surface of the material, the individual data spots can be packed very close together giving a much higher information density for the surface area used than earlier waveguide-based memories can achieve. We believe that this makes our device an interesting and important contender in the next generation of optical memory, especially one that can be used in conjuntion with free-space two-dimensional optical computing protocols\cite{Ambs}. 

\section{Conclusion}
\label{sec:conclusion}

In summary, we have used a custom built optical setup capable of active optical monitoring to probe the time dynamics of the crystallization of GST at 1550 nm incident light. We discovered that there was hardly any difference in the time scales for the crystallization based on pulse size, shape, or power at 1550 nm laser irradiation. We identified four main areas of interest in the crystallization of the GST: an initial nucleation step while the pulse is illuminating the sample that continues until melting occurs, the dropping of the reflectivity as the GST melts and approaches a more disordered state, the diffusion of the heat through the material and a small amount of crystallization indicated by the slowly rising reflectivity, and finally a sharp rise in the reflectivity as the material cools below the crystallization temperature and settles into its final crystallized state. This whole process was seen to take approximately 400 ns. We saw this in the numerical simulations and verified it experimentally.

After verifying the ability of the GST to remain in partially crystallized states, we were able to encode and decode a message written in a 4-bit hexadecimal system, which could be thought of as either a 4-bit single-site memory or a 16-level accumulator. We wrote and read the message "SET/RES" in a 2D multilevel memory application. With tuning of the laser powers, the GST thickness, and the consistency of the laser pulses, we believe we can achieve lower errors and even more levels, making this a viable optical memory.

\begin{acknowledgement}

The authors thank Pengfei Guo, Ryan Laing, and Aditya Sood for their professional input and assistance with the work presented in this manuscript.

\end{acknowledgement}



%

\bibliography{GSTmultiPaper}

\end{document}